\documentclass[10pt,english,11pt,aps,letter,superscriptaddress,jcp]{revtex4}
\usepackage[T1]{fontenc}
\usepackage[latin9]{inputenc}
\usepackage{xcolor}
\usepackage{babel}
\usepackage{verbatim}
\usepackage{amsmath}
\usepackage{amssymb}
\usepackage{graphicx}
\usepackage{esint}
\PassOptionsToPackage{normalem}{ulem}
\usepackage{ulem}

\makeatletter

\@ifundefined{textcolor}{}
{%
 \definecolor{BLACK}{gray}{0}
 \definecolor{WHITE}{gray}{1}
 \definecolor{RED}{rgb}{1,0,0}
 \definecolor{GREEN}{rgb}{0,1,0}
 \definecolor{BLUE}{rgb}{0,0,1}
 \definecolor{CYAN}{cmyk}{1,0,0,0}
 \definecolor{MAGENTA}{cmyk}{0,1,0,0}
 \definecolor{YELLOW}{cmyk}{0,0,1,0}
}

\usepackage{ae,aecompl}

\newcommand{\sfrac}[2]{\mathchoice
  {\kern0em\raise.5ex\hbox{\the\scriptfont0 #1}\kern-.15em/
   \kern-.15em\lower.25ex\hbox{\the\scriptfont0 #2}}
  {\kern0em\raise.5ex\hbox{\the\scriptfont0 #1}\kern-.15em/
   \kern-.15em\lower.25ex\hbox{\the\scriptfont0 #2}}
  {\kern0em\raise.5ex\hbox{\the\scriptscriptfont0 #1}\kern-.2em/
   \kern-.15em\lower.25ex\hbox{\the\scriptscriptfont0 #2}}
  {#1\!/#2}}

\makeatletter
\DeclareMathSizes{\@xipt}{10}{6}{5}
\makeatother

\makeatother

\begin{document}

\title{The Stokes-Einstein Relation at Moderate Schmidt Number}

\author{Florencio Balboa Usabiaga}

\affiliation{Departamento de Física Teórica de la Materia Condensada, Univeridad
Autónoma de Madrid, Madrid 28049, Spain}

\author{Xiaoyi Xie}

\affiliation{Department of Physics, New York University, New York, NY 10012}

\author{Rafael Delgado-Buscalioni}

\affiliation{Departamento de Física Teórica de la Materia Condensada, Univeridad
Autónoma de Madrid, Madrid 28049, Spain}

\author{Aleksandar Donev}

\email{donev@courant.nyu.edu}

\selectlanguage{english}%

\affiliation{Courant Institute of Mathematical Sciences, New York University,
New York, NY 10012}
\begin{abstract}
The Stokes-Einstein relation for the self-diffusion coefficient of
a spherical particle suspended in an incompressible fluid is an asymptotic
result in the limit of large Schmidt number, that is, when momentum
diffuses much faster than the particle. When the Schmidt number is
moderate, which happens in most particle methods for hydrodynamics,
deviations from the Stokes-Einstein prediction are expected. We study
these corrections computationally using a recently-developed minimally-resolved
method for coupling particles to an incompressible fluctuating fluid
in both two and three dimensions. We find that for moderate Schmidt
numbers the diffusion coefficient is reduced relative to the Stokes-Einstein
prediction by an amount inversely proportional to the Schmidt number
in both two and three dimensions. We find, however, that the Einstein
formula is obeyed at all Schmidt numbers, consistent with linear response
theory. The mismatch arises because thermal fluctuations affect the
drag coefficient for a particle due to the nonlinear nature of the
fluid-particle coupling. The numerical data is in good agreement with
an approximate self-consistent theory, which can be used to estimate
finite-Schmidt number corrections in a variety of methods. Our results
indicate that the corrections to the Stokes-Einstein formula come
primarily from the fact that the particle itself diffuses together
with the momentum. Our study separates effects coming from corrections
to no-slip hydrodynamics from those of finite separation of time scales,
allowing for a better understanding of widely observed deviations
from the Stokes-Einstein prediction in particle methods such as molecular
dynamics.
\end{abstract}
\maketitle
\global\long\def\V#1{\boldsymbol{#1}}
\global\long\def\M#1{\boldsymbol{#1}}
\global\long\def\Set#1{\mathbb{#1}}

\global\long\def\D#1{\Delta#1}
\global\long\def\d#1{\delta#1}

\global\long\def\norm#1{\left\Vert #1\right\Vert }
\global\long\def\abs#1{\left|#1\right|}

\global\long\def\grad{\M{\nabla}}
\global\long\def\avv#1{\langle#1\rangle}
\global\long\def\av#1{\left\langle #1\right\rangle }

\global\long\def\P{\mathcal{P}}

\global\long\def\ki{k}
\global\long\def\wi{\omega}

\global\long\def\pRe{\mathrm{Re}_{P}}
 \global\long\def\fRe{\mathrm{Re}_{F}}
 \global\long\def\bu{{\bf u}}
 \global\long\def\bv{{\bf v}}
 \global\long\def\br{{\bf r}}

\section{Introduction}

The self-diffusion coefficient $\chi$ of a tracer particle suspended
in a liquid (for example, a colloidal particle) is a quantity of fundamental
interest, and can be predicted using the well-known Stokes-Einstein
(SE) relation. This famous formula is a combination of two results.
The first, by Einstein, comes from rather general linear response
theory considerations and relates the diffusion coefficient to the
mobility $\mu$, $\chi=k_{B}T\mu$ with $k_{B}T$ the thermal energy
scale, where the mobility is defined by measuring the \emph{average}
steady velocity $\av u$ of the particle in response to a weak applied
force $F$, $\mu=\lim_{F\rightarrow0}\,\av u/F$. The second part
of the SE relation is the Stokes formula for the mobility of a sphere
suspended in a viscous fluid, obtained using standard hydrodynamics.
While there is little reason to doubt the applicability of Einstein's
relation, the Stokes-Einstein relation assumes that the drag force
on a particle is not affected by thermal fluctuations, that is, it
is assumed that the contribution of thermal kicks to the particle
average to zero and the drag is the same as in a deterministic fluid.
As we explain in this work, this is not necessarily so because of
the nonlinear coupling between the \emph{moving} particle and the
fluctuating fluid. 

For a spherical particle of radius $a$ suspended in a three-dimensional
fluid with shear viscosity $\eta$, the SE formula takes the form
\begin{equation}
\chi=\frac{k_{B}T}{\alpha\eta a},\label{eq:SE}
\end{equation}
where $\alpha$ is a coefficient that depends on the boundary conditions
applicable at the surface of the sphere, equal to $6\pi$ for a stick
surface and $4\pi$ for a slip surface \cite{Landau:Fluid}. In fact,
the precise definition of $a$ and $\alpha$ are ambiguous except
in certain limiting cases \cite{StokesEinstein_BCs,StokesEinstein_MD},
and it perhaps best to think of the product $\alpha a=6\pi R_{H}$
as a measure of the effective (stick) \emph{hydrodynamic radius} of
the particle $R_{H}$. It is not difficult to extend the SE relation
to account for rotational diffusion and thus generalize it to more
complicated rigid particle shapes. There is ample evidence and many
theoretical calculations \cite{DiffusionRenormalization_III,BilinearModeCoupling_SE}
that demonstrate that (\ref{eq:SE}) is asymptotically exact for a
rigid sphere that is much larger and much more massive than the solvent
molecules. Remarkably, the SE formula (\ref{eq:SE}) is consistent
with experimental and numerical measurements even for particles that
are comparable in size to the solvent molecules (including a tagged
fluid particle), with values of the hydrodynamic radius that are comparable
to the actual physical size of the particle. Many deviations from
this relation have been observed in particle simulations \cite{StokesEinstein_MD},
however, it is virtually impossible to precisely attribute the cause
of the mismatch because of the large number of violations of the assumptions
that underlie (\ref{eq:SE}). For example, traditional Navier-Stokes
hydrodynamics may break down at the scales of the suspended particle
\cite{GeneralizedHydrodynamics}, or the appropriate boundary conditions
may be different from the traditional no-slip condition \cite{StokesEinstein_BCs}.

Even if one assumes traditional hydrodynamics applies, there are additional
assumptions that enter the SE relation. One of the most important,
yet often overlooked, assumptions, is that the Schmidt number is very
large, $\text{Sc}=\nu/\chi\gg1$, where $\nu=\eta/\rho$ is the kinematic
viscosity of the fluid. Physically, this means that momentum diffuses
much faster than does the particle and the particle motion is viscous-dominated.
In particular, at large Schmidt numbers the fluid velocity quickly
relaxes to the solution of the steady Stokes equation as the particle
barely moves \cite{BrownianParticle_SIBM}. This assumption can safely
be made for realistic liquids. The Schmidt number for molecules in
simple liquids is on the order of $10^{2}-10^{3}$, and for macromolecular
or colloidal solutions it is at least an order of magnitude larger.
However, there are several numerical methods used for simulating the
diffusive motion of particles in flow for which this assumption cannot
be safely made. This is particularly true for particle methods for
hydrodynamics such as smooth particle hydrodynamics (SPH) \cite{Bian2012},
smoothed dissipative particle dynamics (SDPD) \cite{SDPD_Scaling},
stochastic hard-sphere dynamics (SHSD) \cite{SHSD_PRL}, and stochastic
rotation dynamics (SRD) (also called multiparticle collision dynamics
(MPCD) \cite{MPC_LLNS}) \cite{SRD_Review,MPCD_Review}. Achieving
large Schmidt number in these methods requires a prohibitively small
time step (collision frequency) and in many typical simulations $\text{Sc}$
is less than $10$. In several hydrodynamic (non-particle) methods
such as the lattice Boltzmann (LB) method \cite{LB_SoftMatter_Review},
the stochastic immersed boundary method \cite{StochasticImmersedBoundary,SELM},
or the inertial coupling method used here \cite{ISIBM}, $\text{Sc}$
can be varied over a much broader range with comparative ease, however,
$\text{Sc}$ is still limited to moderate values (e.g., $\text{Sc}\sim25-50$)
by computational efficiency considerations due to the large separation
of time scales between the viscous and diffusive dynamics. Practical
heuristics have been developed empirically \cite{LBM_vs_BD_Burkhard,BD_LB_Comparison},
with findings generally consistent with the detailed investigation
we perform in this work.

In this work, we study the deviations of the diffusion coefficient
of a tracer particle immersed in an incompressible Newtonian fluid
from the Stokes-Einstein prediction at small and moderate Schmidt
numbers. We utilize a recently-developed Incompressible Inertial Coupling
Method \cite{ISIBM} method for coupling minimally-resolved particles
with an incompressible fluctuating fluid solver. This allows us to
eliminate compressibility yet consistently include both fluid \emph{and}
particle inertia, as well as the thermal fluctuations responsible
for the diffusive Brownian motion. By changing the fluid viscosity
we are able to control the Schmidt number and thus study whether the
SE formula applies when the particle diffuses comparably fast to momentum.
Furthermore, we can trivially change the dimensionality and study
both two and three dimensional systems. There has been some confusion
in the literature about the applicability of hydrodynamics to two
dimensional systems, and statements to the effect that the SE relation
does not apply in two dimensions have been made \cite{SE_2D}. As
we will demonstrate, these mis-perceptions arise because finite-size
effects diverge in two dimensions \cite{DiffusionRenormalization},
and not because (fluctuating) hydrodynamics fails in two dimensions.
In particular, we will demonstrate that in \emph{finite} two dimensional
systems the SE relation holds for $\text{Sc}\gg1$.

In Section \ref{sec:BlobModel} we describe the formulation of the
minimally-resolved model we use to describe the coupled particle-fluid
system. We first discuss the emergence of the Stokes-Einstein result
in the limit of infinite Schmidt number, and then, in Section \ref{sub:FiniteSc},
we discuss finite Schmidt number corrections. In Section \ref{sec:Results}
we present numerical results for the velocity autocorrelation function
and long-time diffusion coefficient at finite Schmidt numbers, and
then offer some conclusions in Section \ref{sec:Conclusions}.

\section{\label{sec:BlobModel}Brownian Particle Model}

A detailed description of the fluid-particle equations that we employ,
as well as a numerical algorithm to solve them, is given in Ref. \cite{ISIBM}.
Here we briefly summarize the essential features, and then discuss
in more detail the Brownian (overdamped) limit.

Let us consider a particle of physical mass $m$ and size (e.g., radius)
$a$ immersed in a fluid with density $\rho$. The position of the
particle is denoted with $\V q(t)$ and its velocity with $\V u=\dot{\V q}$.
The shape of the particle and its effective interaction with the fluid
is captured through a smooth kernel function $\delta_{a}\left(\V r\right)$
that integrates to unity and whose support is localized in a region
of size $a$. For example, one may choose any one-dimensional ``bell-shaped''
curve $\delta_{a}\left(r\right)$ with half-width of order $a$, such
as a normalized Gaussian with standard deviation $a$ or a symmetric
function of compact support of half-width $a$. This kernel is used
to mediate two crucial operations. First, it is used to transfer (spread)
the force $\V{\lambda}$ exerted on the fluid by the particle to the
fluid. Second, it is used to impose a minimally-resolved form of the
no-slip constraint stating that the velocity of the particle equals
the local velocity of the fluid. We term this diffuse particle a \emph{blob}
for lack of better terminology (in polymer modeling the term bead
is used for the same concept \cite{LB_SoftMatter_Review}, while blob
is used to denote an effective soft particle that includes internal
degrees of freedom)\emph{. }The physical volume of the blob $\D V$
is determined by the shape and width of the kernel function,
\begin{equation}
\D V=\left[\int\delta_{a}^{2}\left(\V r\right)d\V r\right]^{-1}.\label{eq:dV_JS}
\end{equation}

The equations of motion in our Inertial Coupling Method take the form
\cite{ISIBM},
\begin{eqnarray}
\rho\left(\partial_{t}\V v+\V v\cdot\grad\V v\right) & = & -\grad\pi+\eta\grad^{2}\V v+\grad\cdot\left[\left(k_{B}T\eta\right)^{\frac{1}{2}}\left(\M{\mathcal{W}}+\M{\mathcal{W}}^{T}\right)\right]-\V{\lambda}\,\delta_{a}\left(\V q-\V r\right)\label{eq:v_t}\\
\grad\cdot\V v & = & 0\\
m_{e}\dot{\V u} & = & \V F\left(\V q\right)+\V{\lambda}\label{eq:u_t}\\
\V u & = & \int\delta_{a}\left(\V q-\V r\right)\V v\left(\V r,t\right)\, d\V r,\label{eq:no_slip}
\end{eqnarray}
where $\rho$ is the constant fluid density, $\V v\left(\V r,t\right)$
is the incompressible fluid velocity, $\pi\left(\V r,t\right)$ is
the hydrodynamic pressure, $m_{e}$ is the \emph{excess mass} of the
particle relative to the fluid, and $\V F\left(\V q\right)$ is the
external force applied to the particle. Here the stochastic momentum
flux is modeled using a white-noise random Gaussian tensor field $\M{\mathcal{W}}\left(\V r,t\right)$,
that is, a tensor field whose components are independent (space-time)
white noise processes \cite{FluctHydroNonEq_Book}. In this instantaneous
frictionless coupling the total fluid-particle force $\V{\lambda}$
is a Lagrange multiplier that enforces the no-slip constraint (\ref{eq:no_slip}).
In this work we will exclusively consider periodic systems, i.e.,
consider diffusion of the Brownian particle on a torus, in either
two or three dimensions. The equations used here only account for
the hydrodynamic contribution to the diffusion coefficient, and not
any additional molecular (kinetic) contributions \cite{Friction_Diffusion_SE}.
One can include a random slip to account for unresolved molecular
effects, as described in the Appendix of Ref. \cite{ISIBM}, in which
case the additional ``bare'' diffusion coefficient would simply
be added to the hydrodynamic contribution studied here \cite{StokesLaw}.
In principle the equations employed here can be obtained by coarse-graining
the complete microscopic dynamics with the momentum density field
and the position of the particle as the relevant (slow) degrees of
freedom \cite{SELM_Reduction}.

Let us now assume that the nonlinear inertial effects in the momentum
equation can be neglected (this is easy to check in our method by
simply omitting the advective momentum flux term in the implementation),
and that the immersed particle is neutrally-buoyant, $m_{e}=0$. In
this work we will not carefully study the effects of the particle
excess mass $m_{e}$, however, the results presented in Section \ref{sec:Results}
suggest that $m_{e}$ does not affect the long-time diffusive motion.
This suggests that the dominant contribution to the effect we study
here comes from the transient inertia of the fluid (i.e., the term
$\rho\partial_{t}\V v$ in the momentum equation). Under these assumptions,
we get the simpler equations of motion for the fluid-particle system,
\begin{eqnarray}
\rho\partial_{t}\V v+\grad\pi & = & \eta\grad^{2}\V v+\grad\cdot\left[\left(k_{B}T\eta\right)^{\frac{1}{2}}\left(\M{\mathcal{W}}+\M{\mathcal{W}}^{T}\right)\right]+\V F\left(\V q\right)\,\delta_{a}\left(\V q-\V r\right)\label{eq:v_t_Stokes}\\
\V u=\frac{d\V q}{dt} & = & \int\delta_{a}\left(\V q-\V r\right)\V v\left(\V r,t\right)\, d\V r,\nonumber 
\end{eqnarray}
which are much easier to analyze. Note, however, that even though
the fluid equation is linearized, the no-slip constraint is a nonlinear
constraint because of the presence of the particle position in the
argument of the kernel. These semi-linear equations are also the equations
used in the Stochastic Immersed Boundary Method \cite{StochasticImmersedBoundary}
and the closely-related Stochastic Eulerian-Lagrangian Method \cite{SELM}.
We note however that in our numerical calculations we employ the full
equations (\ref{eq:v_t}-\ref{eq:no_slip}), and only use the linearized
equations for theoretical analysis. The numerical method used to solve
the equations relies on a finite-volume staggered discretization of
the fluid equation \cite{LLNS_Staggered}, and on the immersed boundary
method \cite{IBM_PeskinReview} for handling the fluid-particle interaction
\cite{ISIBM}.

The mobility of a blob is easy to evaluate in the \emph{deterministic}
setting (the stochastic setting will be discussed later on). Consider
applying a constant force $\V F$ on the blob and waiting for the
velocity of the particle to reach a steady value. In the steady state
the fluid velocity solves the steady Stokes equation and can be obtained
explicitly,
\[
\V v\left(\V r\right)=\eta^{-1}\int\M G\left(\V r,\V r^{\prime}\right)\V F\left(\V q\right)\delta_{a}\left(\V q-\V r^{\prime}\right)\, d\V r^{\prime},
\]
where $\M G$ is the Green's function (Oseen tensor) for the steady
Stokes equation %
\footnote{For unbounded three-dimensional systems the Oseen tensor is $\V G\left(\V r^{\prime},\V r^{\prime\prime}\right)=\left(8\pi r\right)^{-1}\left(\M I+r^{-2}\V r\otimes\V r\right)$,
where $\V r=\V r^{\prime}-\V r^{\prime\prime}$.%
} with unit viscosity ($\grad\pi=\grad^{2}\V v+\V f$ subject to $\grad\cdot\V v=0$
and appropriate boundary conditions). Note that for periodic boundaries
the integral of $\V f$ over the unit periodic cell of volume $V$
must vanish; we therefore take the definition of $\M G$ to include
subtracting the total applied force $\M F$ as a uniform force density
$-V^{-1}\M F$ on the right hand side of the momentum equation.

The velocity of the blob $\V u=\M{\mu}\V F$ is determined from the
no-slip constraint, giving the mobility tensor (see also Ref. \cite{SIBM_Brownian})
\begin{equation}
\M{\mu}=\eta^{-1}\int\delta_{a}\left(\V q-\V r\right)\M G\left(\V r,\V r^{\prime}\right)\delta_{a}\left(\V q-\V r^{\prime}\right)\, d\V rd\V r^{\prime}.\label{eq:mu_blob}
\end{equation}
For isotropic systems the mobility tensor is a multiple of the identity
tensor, $\M{\mu}=\mu\M I$. In three dimensions, the scalar mobility
$\mu=\left(\alpha\eta a\right)^{-1}$ can be taken to define an effective
hydrodynamic radius of the blob $R_{H}\sim a$ via the relation for
a no-slip rigid sphere, $\mu\eta=\left(6\pi R_{H}\right)^{-1}$. There
are well-known finite-size corrections to the mobility for periodic
systems with a unit cell of volume $L^{d}$ that scale like $L^{-1}$
in three dimensions \cite{Mobility2D_Hasimoto,LB_SoftMatter_Review,DirectForcing_Balboa}.
The hydrodynamic radius of the blobs that we employ in our numerical
implementation has been measured computationally in Refs. \cite{ISIBM,DirectForcing_Balboa}.
The spatial discretization of the fluid/particle equations leads to
a small violation of translational invariance and isotropy and the
mobility tensor is not exactly constant or diagonal but rather depends
on the precise location of the blob relative to the underlying grid
used to solve the fluid equation \cite{ISIBM}. These discretization
artifacts are small, on the order of a couple percent for the three-point
Peskin kernel \cite{StaggeredIBM}, and a fraction of a percent for
the four-point Peskin kernel \cite{IBM_PeskinReview}. It is also
possible to construct discrete kernel functions with even better translational
invariance at the cost of increasing the support of the kernels and
thus the computational cost of the algorithm.

In two dimensions, the mobility of a blob diverges logarithmically
with system size, consistent with the Stokes paradox for flow past
a cylinder in an unbounded domain. For periodic system with a square
unit cell with area $L^{2}$, the logarithmic divergence of the Green's
function for two-dimensional Stokes flow gives %
\footnote{Note that for truly two-dimensional systems $\rho$ has units of $\mbox{kg}/\mbox{m}^{2}$,
unlike three dimensions where it has units $\mbox{kg}/\mbox{m}^{3}$.
This accounts for the difference in units of viscosity $\eta$ in
the Stokes-Einstein relation ($\nu$ has units of $\mbox{m}^{2}/\mbox{s}$
independent of the dimension). %
} 
\begin{equation}
\mu=\left(4\pi\eta\right)^{-1}\ln\frac{L}{\alpha a},\label{eq:mu_2D}
\end{equation}
where the coefficient $\alpha$ depends on the shape of the kernel
$\delta_{a}$ \cite{DiffusionRenormalization}. This is analogous
to the formula for the mobility of a periodic array of no-slip rigid
cylinders of radius $R_{H}$, $\mu=\left(4\pi\eta\right)^{-1}\ln\left[L/\left(3.708\, R_{H}\right)\right]$
\cite{Mobility2D_Hasimoto}, and can be used to define an effective
hydrodynamic radius for a two-dimensional blob. It is important to
note that while the mobility diverges for an infinite system, for
any finite system the mobility is finite and well-defined even in
two dimensions. There is, in fact, no fundamental difference between
two and three dimensions; it is simply the slower decay of the Green's
function in two dimensions that changes the essential physics.

\subsection{\label{sub:BrownianDynamics}Brownian Dynamics Limit}

The short-time motion of particles immersed in the fluid is known
to be very strongly affected by momentum diffusion and by inertial
effects \cite{VACF_Langevin}. Since our method includes inertial
effects, it is able to produce both the correct short-time and long
time features of the Brownian motion of a particle, where ``short''
refers to time scales after sound waves have decayed. This observation
was made for the Stochastic Immersed Boundary Method in Refs. \cite{BrownianParticle_SIBM,SIBM_Brownian},
and in Ref. \cite{ISIBM} we demonstrated that our Inertial Coupling
Method also correctly captures the known physical effects of particle
and fluid inertia on the velocity autocorrelation function (VACF)
\[
\M C(t)=\av{\V u(t)\otimes\V u(0)},
\]
including the effect of particle excess mass on the short time behavior
of the VACF, as well as the long time power-law tail of the VACF.
Here we focus on the long-time diffusive motion of the particles,
where ``long'' means that the motion is observed at a time scale
$\tau$ over which momentum has diffused throughout the domain and
the VACF has decayed to zero, $\tau>t_{L}=L^{2}/\nu$, where $\nu=\eta/\rho$
is the kinematic viscosity.

In the long-time limit, the motion of a single free particle immersed
in a fluid looks like simple Brownian motion with a diffusion coefficient
that can be defined from the long-time mean square displacement of
the particle. More generally, one defines a time-dependent \emph{diffusion
tensor}, either using the mean-square displacement of the particle,
\[
\M{\chi}_{\text{MSD}}(t)=\av{\frac{\left[\V q(t)-\V q(0)\right]\otimes\left[\V q(t)-\V q(0)\right]}{2t}},
\]
or the integral of the VACF,
\[
\M{\chi}_{\text{VACF}}(t)=\int_{0}^{t}\M C(t^{\prime})dt^{\prime}=\frac{d}{dt}\av{\frac{\left[\V q(t)-\V q(0)\right]\otimes\left[\V q(t)-\V q(0)\right]}{2}},
\]
The long-time diffusion tensor is then the asymptotic value
\[
\M{\chi}=\lim_{t\rightarrow\infty}\M{\chi}_{\text{MSD}}(t)=\lim_{t\rightarrow\infty}\M{\chi}_{\text{VACF}}(t).
\]
While we will not demonstrate this here, it can be shown that, for
very large Schmidt numbers, neither sound effects (compressibility)
nor inertial effects affect the diffusion coefficient of a single
particle \cite{VACF_Langevin}. For a single particle, $\M{\chi}$
can be obtained from the mobility tensor $\M{\mu}$ via the Einstein
relation $\M{\chi}=\left(k_{B}T\right)\M{\mu}$. It is important to
note that the diffusive motion of the particle is entirely determined
by its coupling to the fluid and is not an input parameter to our
method. This reflects the physical relationship between fluid velocity
fluctuations (viscosity and temperature) and diffusion coefficient,
as encoded in the Stokes-Einstein relation \cite{StokesLaw}.

The Stokes-Einstein relation can formally be obtained starting from
(\ref{eq:v_t_Stokes}) by taking the \emph{overdamped} limit $\nu=\eta/\rho\rightarrow\infty,$
which can be achieved by either assuming strong viscous friction $\eta\rightarrow\infty$,
or no fluid inertia, $\rho\rightarrow0$. In this limit, momentum
diffuses much faster than does the immersed particles, as measured
by the Schmidt number 
\[
\text{Sc}=\frac{\nu}{\chi}\gg1.
\]
The overdamped limit is the formal limit $\text{Sc}\rightarrow\infty$,
in which the fluid velocity becomes a fast degree of freedom that
can be eliminated adiabatically, while the particle position becomes
a slow degree of freedom \cite{SELM,SELM_Reduction,StokesLaw}. We
emphasize that the Stokes-Einstein relation in a periodic domain of
length $L\gg R_{H}$,
\[
\chi\rightarrow\chi_{\text{SE}}=\left(k_{B}T\right)\begin{cases}
\left(4\pi\eta\right)^{-1}\ln\left[L/\left(3.708\, R_{H}\right)\right] & \mbox{ in two dimensions}\\
\left(6\pi\eta R_{H}\right)^{-1} & \mbox{ in three dimensions}
\end{cases},
\]
can only be justified in the limit $\text{Sc}\rightarrow\infty$.

For a collection of neutrally-buoyant blobs, the multi-dimensional
symmetric positive semi-definite mobility tensor $\M M\left(\V Q\right)=\left\{ \M{\mu}_{ij}\right\} $
(which is a block matrix with blocks of size $d^{2}$, where $d$
is the dimensionality) depends on the positions of all particles $\V Q=\left\{ \V q_{i}\right\} $.
The diagonal block $\M{\mu}_{ii}$ corresponds to the single-particle
mobility in the absence of other particles, while the block corresponding
to particle pair $i$ and $j$ is the inter-particle mobility $\M{\mu}_{ij}$.
For blobs this is given by a generalization of (\ref{eq:mu_blob})
\cite{SIBM_Brownian}, 
\begin{equation}
\M{\mu}_{ij}=\M{\mu}_{ji}=\eta^{-1}\int\delta_{a}\left(\V q_{i}-\V r\right)\M G\left(\V r,\V r^{\prime}\right)\delta_{a}\left(\V q_{j}-\V r^{\prime}\right)\, d\V rd\V r^{\prime}.\label{eq:mu_blob_pair}
\end{equation}
This pairwise hydrodynamic interaction between two blobs was studied
numerically for blobs in Ref. \cite{ISIBM}, and was shown to be closely-related
to the well-known Rotne-Prager-Yamakawa (RPY) tensor used in Brownian
dynamics simulations \cite{BrownianDynamics_DNA,BD_LB_Ladd,LBM_vs_BD_Burkhard}.
In the overdamped limit, the collective diffusion of a collection
of blobs can be described by the equations of Brownian dynamics \cite{SELM,StokesLaw},
\begin{equation}
\frac{d\V Q}{dt}=\M M\V{\mathcal{F}}+\sqrt{2k_{B}T}\,\M M^{\frac{1}{2}}\widetilde{\V{\mathcal{W}}}+k_{B}T\left(\frac{\partial}{\partial\V Q}\cdot\M M\right),\label{eq:brownian_dynamics}
\end{equation}
where $\widetilde{\V{\mathcal{W}}}$ is a collection of independent
white-noise processes and the noise is to be interpreted in the Ito
sense, and $\V{\mathcal{F}}\left(\V Q\right)=\left\{ \V F_{i}\right\} $
are the forces applied on the particles. In this work we focus on
the Brownian motion of a single particle, however, the effect of Schmidt
number that we study here exists in multi-particle systems as well.

It is important to point out that at times longer than the time $t_{L}=L^{2}/\nu$
it takes for momentum to traverse the system length $L$, the VACF
starts decaying exponentially \cite{BrownianParticle_SIBM} and $\M{\chi}_{\text{VACF}}\left(t>t_{L}\right)\approx\mbox{\ensuremath{\M{\chi}}}$.
Therefore, at times $t>t_{L}$ the diffusive motion of an isolated
particle looks like simple Brownian motion and it becomes meaningful
to use the long-time diffusion coefficient. Note however that the
Brownian motions of \emph{multiple} particles are \emph{not} independent
because of the correlations induced by the long-ranged (decaying as
$\ln(r)$ in two dimensions and $r^{-1}$ in three dimensions) hydrodynamic
interactions. The diffusion of a large collection of particles is
therefore subtly but crucially different from that of a collection
of independent Brownian walkers with diffusion coefficient $\chi$
\cite{StokesLaw}.

\subsection{\label{sub:FiniteSc}Finite Schmidt Numbers}

For finite Schmidt numbers, $\text{Sc}=O(1)$, theoretical analysis
is significantly complicated by the fact that the particle moves while
the velocity relaxes through viscous dissipation. We are not aware
of any rigorous results regarding the diffusive motion of even a single
particle, yet alone collective diffusion. The main theoretical approach
in the literature are mode-mode coupling theories \cite{ModeModeCoupling},
which are essentially a perturbative series in the strength of the
thermal fluctuations. One key prediction of these theories is that
the momentum diffusion coefficient $\nu=\eta/\rho$ should be augmented
by the particle diffusion coefficient $\chi$ since the particle diffuses
while the momentum diffuses. At the viscous time scale, $t>t_{\nu}=R_{H}^{2}/\nu$,
conservation of momentum (hydrodynamics) in the fluid introduces a
memory in the motion of the particle and the VACF $C(t)=d^{-1}\av{\V u(t)\cdot\V u(0)}$
decays algebraically rather than exponentially. The standard theory
for the tail of the VACF (long-time behavior) \cite{VACF_Langevin}
implicitly assumes that $\text{Sc}\gg1$, and leads to the conclusion
that for an isolated particle in an infinite fluid asymptotically
$C(t)\approx\left(t/t_{\nu}\right)^{-d/2}\sim\left(\nu t\right)^{-d/2}$.
Self-consistent mode coupling theory predicts that at finite Schmidt
numbers $C(t)\sim\left[\left(\chi+\nu\right)t\right]^{-d/2}$ \cite{VACF_Alder,BilinearModeCoupling_SE}.
This was confirmed to hold for blob particles numerically in Ref.
\cite{ISIBM} for $\text{Sc}\gtrsim2$, with the caveat that $\chi$
was approximated by $\chi_{\text{SE}}$, the prediction of the Stokes-Einstein
relation.

It is not difficult to see that predictions of mode-mode coupling
theories have to be approximate in nature since the diffusion coefficient
of the particle, which is the \emph{result} of the fluid-particle
coupling, is used in the theory as an \emph{input} to predict the
corrections to the overdamped limit. Self-consistent mode-mode coupling
theories are usually heuristic and thus also approximate. Based on
the scaling of the tail of the VACF with Schmidt number, and the fact
that the diffusion coefficient is given by the integral of the VACF,
one may conjecture that, to leading order, the effect of finite Schmidt
number can be captured by the modified Stokes-Einstein formula
\begin{equation}
\chi\left(\nu+\chi\right)=\chi_{\text{SE}}\nu.\label{eq:SE_Bedeaux}
\end{equation}
This relation was proposed as a self-consistent equation for $\chi$
in Refs. \cite{DiffusionRenormalization_I,DiffusionRenormalization_II}.
It is important to note, however, that both the short time \emph{and}
the long time part of the VACF contribute to the diffusion coefficient.
Numerical results in Ref. \cite{ISIBM} indicate that the short-time
VACF does not scale in the same way as the long-time tail, leading
us to question the prediction (\ref{eq:SE_Bedeaux}). In particular,
at very short times no rescaling is required to overlap the VACFs,
and therefore, in principle, to fully overlap the VACFs over both
short and long times, one would need to use an $\text{Sc}$-dependent
non-uniform rescaling of time axes. One therefore expects the integral
of the VACF to be somewhere in-between the Stokes-Einstein relation
(corresponding to the short-time scaling) and (\ref{eq:SE_Bedeaux})
(corresponding to the long-time scaling).

We are not aware of any detailed theory that successfully predicts
the form of the corrections to the simple Stokes-Einstein prediction
(\ref{eq:SE}) for moderate values of $\text{Sc}$ (see Ref. \cite{BilinearModeCoupling_SE}
for one attempt). In Ref. \cite{DiffusionRenormalization} a heuristic
self-consistent theory of diffusion is proposed. The starting point
is renormalization theory for the effective diffusion coefficient
in the advection-diffusion equation for the concentration of a large
number of passive tracers %
\footnote{Note that freely diffusing neutrally-buoyant blobs are passive tracers
since they simply follow the fluid but do not affect it.%
}. The self-consistent theory suggests that in both two and three dimensions
(c.f. Eqs. (29) and (30) in Ref. \cite{DiffusionRenormalization}
in the case of no bare diffusion)

\begin{equation}
\chi=\nu\left[1+2\frac{\chi_{\text{SE}}}{\nu}\right]^{1/2}-\nu=\nu\left[\left(1+\frac{2}{\text{Sc}^{\text{SE}}}\right)^{\frac{1}{2}}-1\right],\label{eq:chi_eff_2D_SC}
\end{equation}
where we have defined a predicted Schmidt number $\text{Sc}^{\text{SE}}=\nu/\chi_{\text{SE}}$
(which is input to our simulations rather than output as is $\text{Sc}$).
This prediction for the self-diffusion coefficient is the solution
to the self-consistent equation
\begin{equation}
\chi\left(\nu+\frac{\chi}{2}\right)=\chi_{\text{SE}}\nu,\label{eq:SE_self_consistent}
\end{equation}
which differs from (\ref{eq:SE_Bedeaux}) in the prefactor of $1/2$.
In the next section, we will compare the two predictions (\ref{eq:SE_Bedeaux})
and (\ref{eq:SE_self_consistent}) for the leading-order correction
to Stokes-Einstein's relation with numerical results. Note that we
only expect this type of relation to predict the leading-order corrections
(proportional to $\text{Sc}^{-1}$) to the Stokes-Einstein relation
and not the complete dependence on $\text{Sc}$. At very small Schmidt
numbers there would likely be important contributions from higher-order
(e.g., proportional to $\chi^{2}/\nu$) correction terms inside the
parenthesis in (\ref{eq:SE_self_consistent}).

Mode-mode and renormalization theories, on which both (\ref{eq:SE_Bedeaux})
and (\ref{eq:SE_self_consistent}) are based, are perturbative theories
typically truncated at terms quadratic in the strength of the fluctuations.
In the context of infinite (bulk) systems, a systematic perturbative
theory that accounts for corrections of order higher than quadratic
in the fluctuations has been discussed in Refs. \cite{DiffusionRenormalization_II,TrilinearModeCoupling}.
In three dimensions, the conclusion of such studies has been that
the higher-order terms do not make a dramatic contribution. This can
be attributed to the fact that large-scale modes of the fluctuating
velocity make a negligible contribution to the diffusive dynamics
(this is directly related to the fact that the $t^{-3/2}$ tail of
the VACF is integrable, see Ref. \cite{DiffusionRenormalization}
for additional discussion).

In two dimensions, however, the logarithmic divergence of $\chi_{\text{SE}}$
with system size is a sign of the growing contribution of large-scale
modes. In fact, the Schmidt number will become arbitrarily small for
sufficiently large systems (keeping viscosity fixed) and the Stokes-Einstein
relation will be strongly violated. In this case, however, the infinite-time
diffusion coefficient $\chi$ does not describe the diffusive dynamics
even at macroscopic length (e.g., system size $L$) and time scales
($L^{2}/\chi$). Specifically, the slowly-decaying $t^{-1}$ tail
of the VACF means that the diffusive motion has correlations over
macroscopic times and does not resemble simple Brownian motion at
relevant scales. Instead, one needs to consider the time-dependent
diffusion coefficients $\chi_{\text{MSD}}(t)$ or $\chi_{\text{VACF}}(t)$.
Several calculations \cite{VACF_2D_SelfConsistentMC,TrilinearModeCoupling,SCModeModeCoupling2D}
and numerical simulations \cite{VACF_2Divergence,VACF_2D_HS} suggest
that including higher-order terms changes the decay of the tail of
the VACF in two dimensions. Specifically, it has been predicted that
the self-consistent power-law decay for the VACF is faster than $t^{-1}$,
$C(t)\sim\left(t\sqrt{\ln t}\right)^{-1}$. Hydrodynamics-based methods
such as the method we use here (see also Ref. \cite{VACF_2Divergence}
for a study using a lattice gas model) are better for studying this
very long time decay of the VACF than are particle methods, since
larger systems and longer time scales become computationally-feasible.
Nevertheless, because of the slow logarithmic growth, extremely large
systems are required to see any effects of higher-order corrections.
To see this, let us assume Stokes-Einstein's formula holds and estimate
the system size when
\[
\text{Sc}\approx\text{Sc}^{\text{SE}}=\frac{\nu}{\chi_{\text{SE}}}=\frac{4\pi\rho\nu^{2}}{k_{B}T\ln\left[L/\left(4\, R_{H}\right)\right]}<1.
\]
For a hard-disk fluid, we can estimate the viscosity $\eta$ using
simple Enskog kinetic theory \cite{Enskog_2DHS}, which has been found
to be quite accurate even at high densities \cite{Hard_Disks_Transport}.
For a hard-disk fluid at packing density (fraction) $\phi=\left(\rho/m\right)\left(\pi\sigma^{2}/4\right)=0.6$
(which is close to the liquid-solid transition), Enskog theory predicts
\[
\eta\approx2\frac{\sqrt{mk_{B}T}}{\sigma},
\]
where $\sigma\approx2R_{H}$ denotes the particle diameter. This gives
that the system size required to get Schmidt number close to unity,
\[
\text{Sc}^{\text{SE}}\approx\frac{65}{\ln\left[L/\left(4\, R_{H}\right)\right]}<1\quad\Rightarrow\quad L>1.5\cdot10^{29}\, R_{H}.
\]
Reaching such a staggering system size is not feasible with any numerical
method, and therefore in this work we focus on the more practically-relevant
case of finite system size at moderately large Schmidt numbers.

In addition to the nonlinearity coming from the fact that at moderate
Schmidt numbers the particle moves while momentum diffuses %
\footnote{In our model, (\ref{eq:no_slip}) has a non-linear coupling between
fluid and particle velocities. In more traditional models the no-slip
boundary condition needs to be applied on a moving surface, which
is hard and not done in most models, see for example the assumptions
made in Ref. \cite{VACF_Langevin}.%
}, additional nonlinearities arise because of the presence of the advective
term $\rho\V v\cdot\grad\V v$ in the momentum equation. If we define
a ``thermal'' Reynolds number based on the equilibrium magnitude
of the particle thermal velocity fluctuations \cite{ISIBM}
\[
\av{u^{2}}^{\frac{1}{2}}\approx\sqrt{\frac{k_{B}T}{\rho R_{H}^{d}}},
\]
in three dimensions we can estimate
\[
\text{Re}=\frac{\av{u^{2}}^{\frac{1}{2}}R_{H}}{\nu}\approx\sqrt{\frac{k_{B}T}{\nu\eta R_{H}}}\sim\sqrt{\frac{\chi_{\text{SE}}}{\nu}}=\left(\text{Sc}^{\text{SE}}\right)^{-\frac{1}{2}}.
\]
A similar result applies in two dimensions as well, except for an
additional factor of $\ln^{\frac{1}{2}}\left[L/\left(4\, R_{H}\right)\right]$.
This shows that the thermal Reynolds number becomes $O(1)$ when $\text{Sc}^{\text{SE}}\sim1$,
and in principle nonlinear effects arising from advection could arise.
By contrast, at large Schmidt numbers the thermal Reynolds number
is small and advective nonlinearities are expected to be negligible.
While in realistic fluids, including particle simulations, advective
nonlinearities are always present, in our numerical method we can
simply turn off the term $\rho\V v\cdot\grad\V v$ to assess its importance.
Numerical experiments indicate that nonlinear inertial effects play
a minimal role, not noticeable within statistical accuracy for reasonable
Schmidt numbers. Therefore, it appears appropriate to simply linearize
the velocity equation, as is often done in the literature with somewhat
hand-waving justifications. Similarly, we find that the excess inertia
of the particles does not affect the long-time diffusion coefficient,
consistent with traditional derivations based on linearized theory
\cite{VACF_Langevin}. We do not attempt to provide rigorous justification
for these observations in this work.

\section{\label{sec:Results}Results}

In this section we present numerical results obtained using the computational
algorithm for solving (\ref{eq:v_t}-\ref{eq:no_slip}) developed
in Ref. \cite{ISIBM}. In Ref. \cite{ISIBM} some of us presented
results concerning the short-time behavior of the VACF $C(t)$ of
a single blob particle in a three-dimensional periodic domain of length
$L$, including a comparison to the case of a blob suspended in a
compressible fluid. Excellent agreement with theoretical predictions
for the variance of velocity (i.e., $C(t=0)$) was obtained, and the
$t^{-3/2}$ power-law behavior at long times $L^{2}/\nu>t>t_{\nu}=R_{H}^{2}/\nu$
was confirmed. Here we briefly examine the VACF in two dimensions,
and then focus on the self-diffusion of a blob in both two and three
dimensions. Note that in Ref. \cite{ISIBM} it is demonstrated that
the method correctly reproduces the static structure of a suspension
of many interacting blobs. Also note that the presence of multiple
interacting blobs affects the dynamics in a nontrivial way, and changes
both the short-time VACF (via multi-particle inertial effects) and
also the self-diffusion coefficient. We do not address multi-particle
suspensions in this work.

In the majority of the simulations we use the three-point discrete
kernel function of Roma and Peskin \cite{StaggeredIBM,DirectForcing_Balboa}
to discretize the kernel $\delta_{a}$. By using the Peskin four-point
kernel \cite{IBM_PeskinReview} instead of the three-point discrete
kernel function the translational invariance of the spatial discretization
can be improved, however, at a potentially significant increase in
computational cost, particularly in three dimensions. The effective
hydrodynamic radius $R_{H}$ for a given discrete kernel function
can be obtained easily from the deterministic mobility tensor $\V{\mu}_{\text{det}}$,
given by a discrete equivalent of (\ref{eq:mu_blob}). The deterministic
mobility tensor can be obtained by turning off fluctuations, applying
a unit force along each of the coordinate directions in turn, solving
the spatially-discretized steady Stokes equation, and then calculating
the resulting velocity of the particle. After accounting for finite-size
effects due to the finite length of the periodic box, in three dimensions
we numerically estimate \cite{DirectForcing_Balboa,ISIBM} the effective
hydrodynamic radius to be $R_{H}^{3pt}=\left(0.91\pm0.01\right)h$
the three-point kernel %
\footnote{The small variation around the average value comes because the spatial
discretization is only approximately translationally invariant.%
}, where $h$ is the grid spacing, and $R_{H}^{4pt}=\left(1.255\pm0.005\right)h$
for the 4pt kernel \cite{ReactiveBlobs}. In two dimensions, the effective
(rigid disk) hydrodynamic radii are estimated to be $R_{H}^{3pt}=\left(0.72\pm0.01\right)h$
and $R_{H}^{4pt}=\left(1.04\pm0.005\right)h$. Note that the spatial
discretization we use is not perfectly translationally invariant and
there is a small variation of $R_{H}$ (quoted above as an error bar)
as the particle moves relative to the underlying fixed fluid grid
\cite{ISIBM,ReactiveBlobs}.

We use a relatively small grid of $32^{2}$ cells in two dimensions
and $32^{3}$ cells in three dimensions in order to be able to perform
sufficiently long runs even with the larger Schmidt numbers. In two
dimensions we use a neutrally-buoyant particle ($m_{e}=0$), while
in three dimensions we use a particle twice denser than the surrounding
fluid ($m_{e}=\rho\D V$), in order to confirm that the excess mass
(density) does not (significantly) affect the conclusions of our investigations.
The advantage of using neutrally-buoyant blobs is that they are passive
tracers (they do not affect the velocity equation), and therefore
one can use multiple tracers in a single simulation in order to improve
the statistical accuracy. This is useful in two dimensions, where
the VACF has a slowly-decaying tail and therefore long runs are required
to study the long-time diffusive motion of the particle. In all cases
we have confirmed that the time step size $\D t$ is sufficiently
small by comparing with a simulation using a twice smaller time step
size. A detailed discussion and numerical results concerning the accuracy
of the scheme as a function of $\D t$ are given in Ref. \cite{ISIBM}.
We varied the viscosity in order to change the Schmidt number, but
as explained earlier, the Schmidt number is the only relevant dimensionless
number so one can equivalently change the temperature. Since the actual
Schmidt number is an output rather than an input to our calculation,
we cannot calculate the Schmidt number a priori. Therefore, in this
section we estimate the true Schmidt number with $\text{Sc}\approx\text{Sc}^{\text{SE}}=\nu/\chi_{\text{SE}}$
for the purposes of captions and axes labels, which is a good approximation
for moderate and large Schmidt numbers.

During each simulation a discrete trajectory of the particle $\left\{ \V q\left(0\right),\,\V q\left(\D t\right),\,\V q\left(2\D t\right),\dots\right\} $
is recorded. From this data, we calculate an apparent (discrete) velocity
$\tilde{\V u}_{k}=\left(\V q(k+1)-\V q(k)\right)/\D t,$ $k=1,2,\dots$,
which for finite $\D t$ is in general different from the velocity
of the particle $\V u\left(k\D t\right)$ calculated by the numerical
scheme. We then obtain the discrete VACF $C_{k}=d^{-1}\av{\left(\tilde{\V u}_{k^{\prime}}\right)\cdot\left(\tilde{\V u}_{k^{\prime}+k}\right)}$
efficiently using a fast Fourier Transform of the apparent velocity.
It is not hard to show that the time-dependent diffusion coefficients
at the discrete time points in time can be obtained in linear time
using
\begin{eqnarray}
\chi_{\text{MSD}}(k\D t) & = & \frac{\D t}{2}C_{0}+\D t\left[\sum_{i=1}^{k-1}C_{i}-\frac{1}{k}\sum_{i=1}^{k-1}i\cdot C_{i}\right]\label{eq:chi_MSD_num}\\
\chi_{\text{VACF}}(k\D t) & \approx & \frac{\D t}{2}C_{0}+\D t\,\sum_{i=1}^{k-1}C_{i}.\label{eq:chi_VACF_num}
\end{eqnarray}

\subsection{\label{sub:VACF_short}VACF}

\begin{figure}
\centering{}\includegraphics[width=0.75\columnwidth]{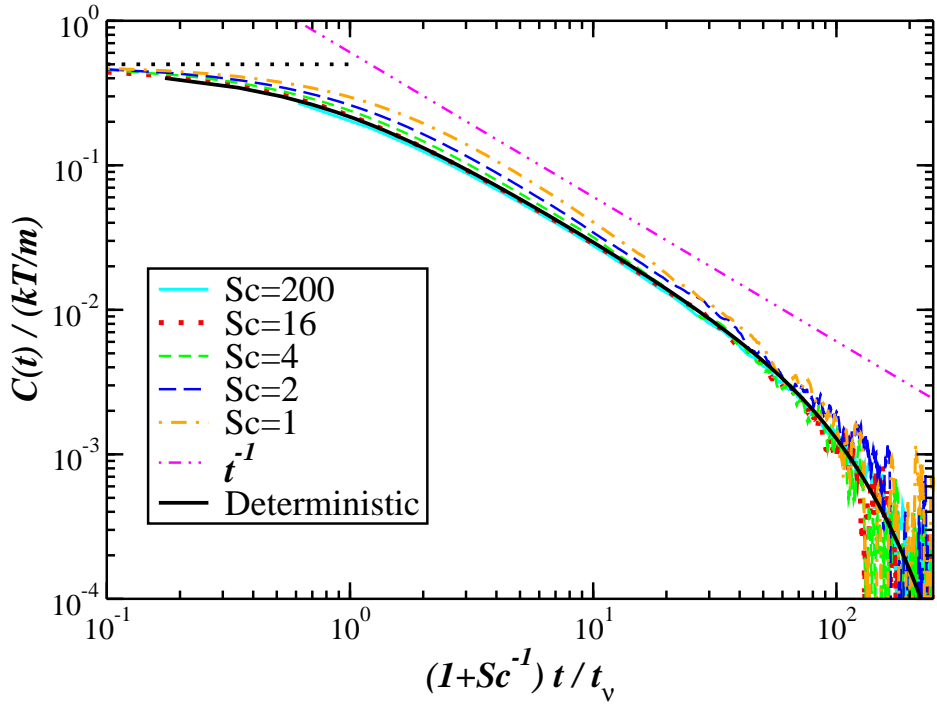}
\caption{\label{fig:VACF}Velocity autocorrelation function (VACF) of a single
neutrally-buoyant blob (three-point kernel) in two dimensions. A deterministic
calculation, corresponding to the limit $\text{Sc}\rightarrow\infty$,
is also shown. The expected value at the origin is shown with a dotted
line. The scaling of the time axes is chosen to collapse the different
curves at long times. Note that the time $t_{L}=L^{2}/\nu\approx10^{3}t_{\nu}$
is off the scale of the time axis.}
\end{figure}

In Fig. \ref{fig:VACF} we show the VACF $C(t)=d^{-1}\av{\V u(t)\cdot\V u(0)}$
for a neutrally-buoyant ($m_{e}=0$) blob in two dimensions, for several
Schmidt numbers. The theoretical variance of the velocity of a neutrally-buoyant
particle immersed in an incompressible two dimensional fluid gives
$C(t=0)=(d-1)k_{B}T/(d\, m)=k_{B}T/\left(2\rho\D V\right)$, where
$\D V$ is the volume of the blob given in Eq. (\ref{eq:dV_JS}).
We see from the figure that the numerical curves are in excellent
agreement with this value, confirming that the numerical method correctly
captures the thermal particle velocity fluctuations.

The standard theory for the tail of the VACF (long-time behavior)
\cite{VACF_Langevin} implicitly assumes that $\text{Sc}\gg1$, and
predicts that the long-time VACF has a power law decay $C(t)\sim\left(\nu t\right)^{-1}$.
As discussed in Section \ref{sub:FiniteSc}, accounting for a finite
Schmidt number leads to a predicted decay $C(t)\sim\left[\left(\nu+\chi\right)t\right]^{-1}$.
This means that the tails of the VACFs for different $\text{Sc}$
values can be collapsed on one master curve if we plot them as a function
of $\left(1+\text{Sc}^{-1}\right)\left(t/t_{\nu}\right)$. This was
confirmed in Ref. \cite{ISIBM} in three dimensions, and it is confirmed
in two dimensions in Fig. \ref{fig:VACF}. The rescaling is not perfect
at shorter times, especially for small Schmidt numbers, however, we
note that the collapse is significantly poorer if one plots the VACF
as a function of just $t/t_{\nu}$.

Our numerical method becomes very inefficient as the Schmidt number
becomes very large due to the large separation of scales between the
momentum and particle diffusion. We will present modifications of
the numerical method to handle the limit $\text{Sc}\rightarrow\infty$
in future work. The VACF in the limit $\text{Sc}\rightarrow\infty$
can, however, easily be obtained by a \emph{deterministic} calculation.
We simply give the particle a small momentum kick $\D p=F\D t$ during
the first time step, and then the deterministic algorithm is used
to track the subsequent decay of the velocity $u(t)$, which gives
the VACF in the limit after a suitable rescaling. In particular, the
time-dependent diffusion coefficient in the limit $\text{Sc}\rightarrow\infty$
is given by
\[
\lim_{\text{Sc}\rightarrow\infty}\chi_{\text{VACF}}(t)=\frac{k_{B}T}{\D p}\int_{0}^{t}u(t^{\prime})dt^{\prime}.
\]
In Fig. \ref{fig:VACF} we show also the result of such a deterministic
calculation. Due to lack of statistical noise, the deterministic VACF
shows very clearly the transition from a power-law to an exponential
tail at long times. It is also clearly seen that the VACF at large
Schmidt numbers closely matches the deterministic one, as expected.
Note that, in principle, the deterministic VACF can be calculated
analytically using discrete Fourier Transform techniques \cite{BrownianParticle_SIBM}.

\subsection{\label{sub:MSD}Diffusion Coefficient}

The long-time diffusion coefficient can be obtained either from the
limiting value of $\chi_{\text{MSD}}(t)$ or $\chi_{\text{VACF}}(t)$.
It is preferable to use the discrete integral of the (discrete) VACF
rather than the mean square displacement because $\chi_{\text{VACF}}$
converges faster to the asymptotic (long-time) value %
\footnote{This can be seen from the fact that the last term in (\ref{eq:chi_MSD_num})
does not appear in (\ref{eq:chi_VACF_num}).%
}, and thus has smaller statistical error. Because the position of
the particle is a more fundamental quantity (in particular, in the
limit $\text{Sc}\rightarrow\infty$ it becomes the only relevant variable),
the mean square displacement has a more direct physical interpretation,
and is directly obtained from the discrete particle trajectory. We
therefore use $\chi_{\text{MSD}}(t)$ to illustrate the time dependence
of the diffusion coefficient, and use $\chi_{\text{VACF}}(t)$ to
obtain the long-time diffusion coefficient $\chi$.

\begin{figure*}
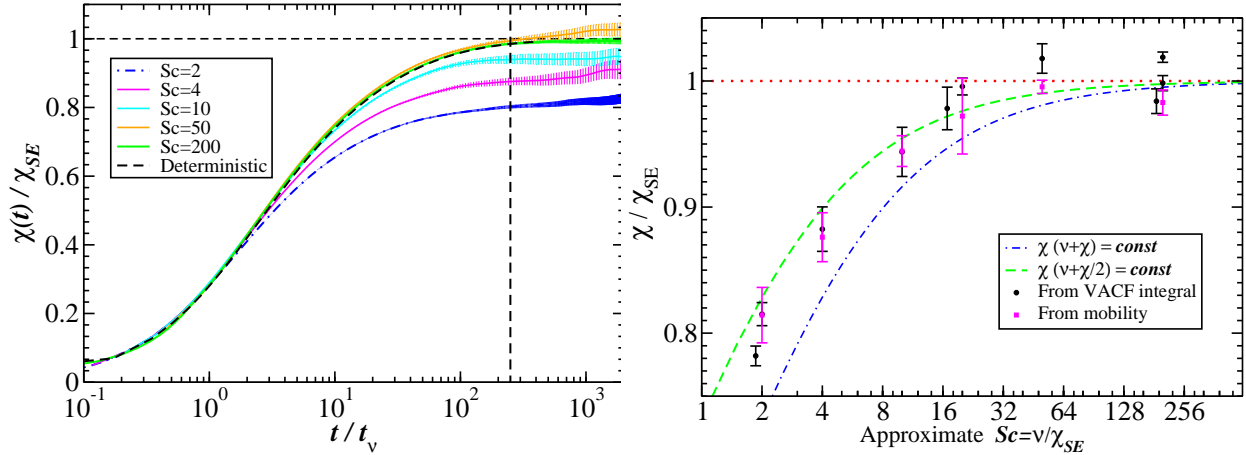

\centering{}\includegraphics[width=0.5\columnwidth]{16_home_donev_Papers_ISIBM_graphics_InertialIncomp_MSD_Schmidt.eps}\includegraphics[width=0.5\columnwidth]{17_home_donev_Papers_ISIBM_graphics_InertialIncomp_diffusion.eps}
\caption{\label{fig:MSD-3D} (\emph{Left panel}) The ratio of the time-dependent
diffusion coefficient $\chi_{\text{MSD}}(t)$ to the long-time diffusion
coefficient estimate $\chi_{\text{SE}}$ given by the Stokes-Einstein
formula, for a three dimensional blob (three-point kernel). The statistical
error estimates are shown as vertical bars, and the vertical line
shows the time $\tau=L^{2}/\left(4\nu\right)$ at which we estimate
the long-time diffusion coefficient $\chi\approx\chi_{\text{VACF}}\left(\tau\right)$.
(\emph{Right panel}) Comparison between the numerical estimate (symbols)
for the long-time diffusion coefficient and the Stokes-Einstein prediction
for several Schmidt numbers. The predictions of the self-consistent
theories (\ref{eq:SE_Bedeaux}) and (\ref{eq:SE_self_consistent})
are shown for comparison (dashed lines). The Einstein prediction $\chi=\left(k_{B}T\right)\mu$
based on direct measurements of the mobility (\ref{eq:Stokes_Einstein_true})
are shown for several Schmidt numbers.}
\end{figure*}

In the left panel of Fig. \ref{fig:MSD-3D} we show the time-dependent
diffusion coefficient $\chi_{\text{MSD}}(t)$ for a single particle
in a three dimensional periodic domain at several Schmidt numbers,
including the limit $\text{Sc}\rightarrow\infty$ as obtained from
a deterministic simulation. We see that at short times we obtain collapse
of all of the curves just by scaling the time axes by $t_{\nu}=R_{H}^{2}/\nu$,
however, at long times the curves do not collapse. In particular,
the diffusion coefficient is lower than the Stokes-Einstein prediction
for the smaller Schmidt numbers. In order to obtain an estimate of
the long-time diffusion coefficient that is essentially converged
to the asymptotic value, while at the same time minimizing the statistical
errors, we estimate $\chi\approx\chi_{\text{VACF}}(L^{2}/\left(4\nu\right))$.
In the right panel of Fig. \ref{fig:MSD-3D} we show the estimated
$\chi/\chi_{\text{SE}}$ as a function of the approximate Schmidt
number $\text{Sc}^{\text{SE}}$. The numerical data is compared to
the two self-consistent theories, (\ref{eq:SE_Bedeaux}) and (\ref{eq:SE_self_consistent}).
We see good agreement of the numerical data with (\ref{eq:SE_self_consistent})
to within the statistical accuracy.

Our numerical results clearly demonstrate that the Stokes-Einstein
relation does not apply at finite Schmidt numbers. On the other hand,
the Einstein formula, which is a linear response relationship between
the diffusion coefficient and the mobility, $\chi=\left(k_{B}T\right)\mu$,
is quite general, and follows from straightforward statistical mechanics
arguments \cite{FluctuationDissipation_Kubo}. However, one must be
careful here in defining the mobility $\mu$. Mobility should be defined
as the coefficient of proportionality between a small applied force
$F_{0}$ on the particle (with the opposite force applied uniformly
to the fluid to prevent center-of-mass motion) and the \emph{average}
velocity of the particle,
\begin{equation}
\mu=\lim_{F_{0}\rightarrow0}\frac{\av{\V u}}{\V F_{0}},\label{eq:Stokes_Einstein_true}
\end{equation}
in the \emph{presence} of thermal fluctuations. Here the average is
taken over time, or, equivalently, over a nonequilibrium but steady-state
ensemble that is a weak perturbation of the equilibrium ensemble.
Because of the nonlinearity of the fluid-particle equations, thermal
fluctuations \emph{can }affect average values. Therefore, we should
not expect, in general, that the mobility measured at finite temperature
is the same as the deterministic ($k_{B}T=0$) mobility. Indeed, in
the right panel of Fig. \ref{fig:MSD-3D} we show the predictions
of (\ref{eq:Stokes_Einstein_true}) with $\av{\V u}$ measured numerically
under a small applied force (to ensure the system remains in the linear
response regime). These predictions are in agreement with the direct
measurements of the diffusion coefficient from the integral of the
velocity autocorrelation function, as predicted by linear response
theory. We therefore observe no violation of any known physical principles,
so long as we recognize the fact that fluctuations cannot just be
turned off. More precisely, at finite value of $k_{B}T$ the fluctuations
do not necessarily average out, and can, in fact, affect the dynamics
of macroscopic or deterministic observables.

\begin{figure*}
\centering{}\includegraphics[width=0.5\columnwidth]{18_home_donev_Papers_ISIBM_graphics_BlobMSD_2D.eps}\includegraphics[width=0.5\columnwidth]{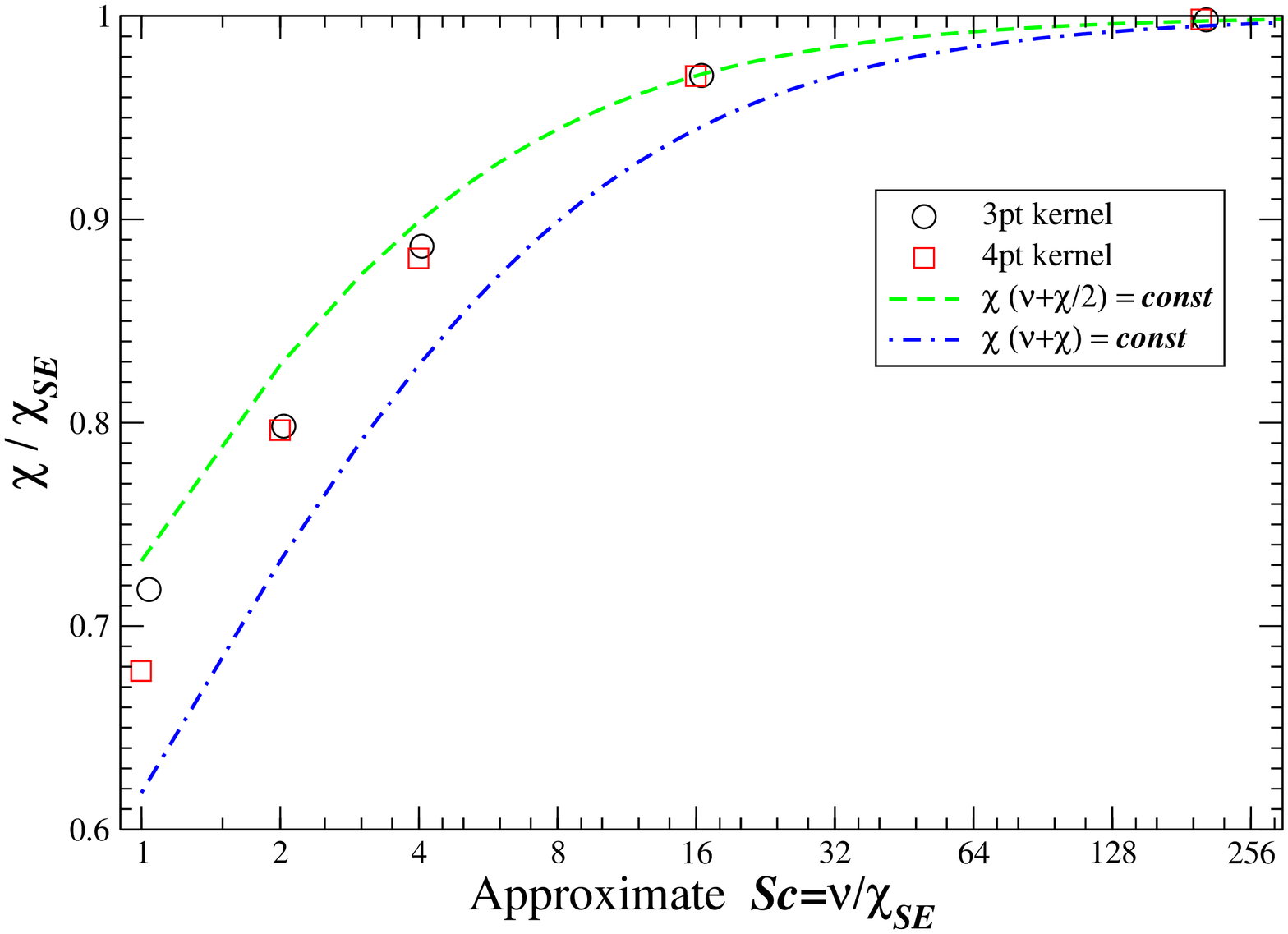}
\caption{\label{fig:MSD-2D} The same panels as Fig. \ref{fig:MSD-3D} but
this time for a two dimensional blob. The four-point kernel was used
for the left panel, and both the three and four-point kernels are
employed for the right panel.}
\end{figure*}

In Fig. \ref{fig:MSD-2D} we confirm that the same conclusions apply
to finite two-dimensional systems. In order to verify that the results
are not affected by the specific choice of the kernel function, here
we also use the four-point Peskin kernel \cite{IBM_PeskinReview}.
The results in the right panel of Fig. \ref{fig:MSD-2D} show that
dimensionless relations are independent of the kernel function. This
leads us to believe that (\ref{eq:SE_self_consistent}) is a rather
general relation that correctly captures the leading-order corrections
to the Stokes-Einstein relation at finite Schmidt numbers in both
two and three dimensions.

\section{\label{sec:Conclusions}Conclusions}

We presented a simple theory for the effect of Schmidt number $\text{Sc}=\nu/\chi$
on the long-time diffusion coefficient of a particle immersed in a
fluid. The approximate self-consistent calculation suggests that the
long-time diffusion coefficient of the particle can be determined
from the condition $\chi\left(\nu+\chi/2\right)=\chi_{\text{SE}}\nu$.
Here $\chi_{\text{SE}}$ is the prediction of the Stokes-Einstein
relation, $\chi_{\text{SE}}=k_{B}T\mu_{det}$, where $\mu_{\text{det}}$
is the deterministic mobility of the particle, that is, the linear
response to an applied force in the \emph{absence} of thermal fluctuations.
In the limit of large Schmidt numbers the Stokes-Einstein relation
is obeyed, as expected, but the correction can be substantial at small
Schmidt numbers and is also measurable at moderate Schmidt numbers.
By contrast, the Einstein relation $\chi=k_{B}T\mu$ is always obeyed,
as long as the mobility is defined by the linear response in the \emph{presence}
of fluctuations. The discrepancy between $\mu$ and $\mu_{\text{det}}$
stems from the nonlinear fluid-particle coupling, which becomes important
at finite Schmidt numbers.

Our numerical results are in good agreement with the self-consistent
theory in both two and three dimensions. In two dimensions, for finite
systems, Stokes-Einstein's relation holds, but the Stokes-Einstein
diffusion coefficient diverges logarithmically with system size. At
the same time, however, as the system size diverges the Schmidt number
decreases and the deviations from Stokes-Einstein's prediction become
stronger. The truly asymptotic behavior of a particle in an unbounded
two-dimensional fluid (e.g., a tagged hard disk in a hard-disk fluid)
remains an open question, albeit one of mostly academic interest since
the Schmidt number is still rather large for system sizes feasible
in present-day molecular dynamics.

While the above results elucidate some important fundamental questions
about the physical importance of fluctuations, it should be observed
that for realistic fluids (rather than coarse-grained fluid models
\cite{MPC_LLNS}) small $\text{Sc}$ implies that the size of the
blob particle becomes comparable to the length scale at which hydrodynamics
breaks down (molecular diameter for liquids, or mean free path for
gases). The physical fidelity of the blob model and the fluctuating
hydrodynamics equations \cite{MPC_LLNS} used here should therefore
be questioned at small $\text{Sc}$. In Ref. \cite{StokesEinstein_MD}
the validity of the Stokes-Einstein relation is studied using molecular
dynamics. In that study the smallest Schmidt number is on the order
of 15, and the corrections studied here are significant. Nevertheless,
in the case of a true molecular fluid many other effects enter the
picture as well, for example, the (in)validity of the no-slip condition
with a fixed particle radius. Therefore, one should not try to explain
the data in Ref. \cite{StokesEinstein_MD} using our model. In fact,
the deviations from the SE prediction measured in Ref. \cite{StokesEinstein_MD}
have a comparable magnitude but the \emph{opposite} sign from the
correction predicted here. This means that, when one takes into account
that hydrodynamic theory should include the finite Schmidt-number
correction studied here, the deviations between the MD simulation
and hydrodynamic predictions are twice larger than reported in Ref.
\cite{StokesEinstein_MD}. A similar ``cancellation-of-errors''
that leads to a better agreement with hydrodynamic theory than truthfully
present is discussed in Ref. \cite{StokesEinstein_BCs}. 

Despite its somewhat limited practical utility, our model allows us
to separate hydrodynamic from molecular effects, and to test the predictions
of self-consistent renormalization theory \cite{DiffusionRenormalization}.
More importantly, our results are quite relevant to the interpretation
of results obtained using widely-used particle methods for fluid dynamics
of complex fluids, e.g., MD, (S)DPD, MPCD/SRD, SHSD, and even Lattice
Boltzmann schemes. In some of these methods, as they are typically
used, the Schmidt number can be on the order of unity, and in all
of them it is rarely larger than 50. This is to be contrasted with
realistic liquids (e.g., water at room conditions) where the Schmidt
number for tagged particles exceeds 100 and is often larger than 1000.
In order to avoid the types of unphysical deviations from the Stokes-Einstein
relation as we described here, one must use methods that take the
limit $\text{Sc}\rightarrow\infty$ and use a fluctuating \emph{steady}
Stokes equation for the fluid (e.g., Brownian and Stokesian dynamics).
In the future we will develop modifications of our method that achieves
this limiting behavior. Note that increasing $\text{Sc}$ is not possible
in traditional particle methods or Lattice Boltzmann methods because
they (by design) avoid the solution of Poisson equations for the pressure,
as is necessary in the steady Stokes limit.

The primary cause of the correction to the SE prediction at finite
$\text{Sc}$ is the fact that the particle diffuses with effective
diffusion coefficient $\chi$ \emph{while} the momentum diffuses,
rather than nonlinear inertial effects. This leads us to suspect that
the effect studied here is independent of the details of the microscopic
dynamics giving rise to the particle diffusion, and the same self-consistent
formula should apply, for example, in Lattice Boltzmann methods with
a frictional fluid-particle coupling \cite{LB_SoftMatter_Review}.
Note that in Stokes frictional coupling, and likely also in molecular
(particle) simulations, there is an additional non-hydrodynamic or
``bare'' contribution to the diffusion coefficient $\chi_{0}=k_{B}T/\gamma$
coming from the dissipative friction coefficient $\gamma$ between
the particle and the fluid \cite{Friction_Diffusion_SE,LB_SoftMatter_Review}.
A way to estimate such bare friction in molecules, from the autocorrelation
of molecular forces in restraint dynamics, has been proposed in Ref.
\cite{MoriZwanzig_ConstrainedMD}. The bare diffusion coefficient
$\chi_{0}$ should be added to the (Schmidt-number dependent) hydrodynamic
contribution $\chi_{H}=\chi-\chi_{0}$ studied in this work. In both
two and three dimensions, the self-consistent theory proposed in Ref.
\cite{DiffusionRenormalization} gives the prediction
\[
\chi=\chi_{0}+\left[1+\frac{\chi+\chi_{0}}{2\nu}\right]^{-1}\chi_{\text{SE}},
\]
which can be written as a self-consistent equation for $\chi_{H}$,
\[
\chi_{H}\left(\nu+\chi_{0}+\frac{\chi_{H}}{2}\right)=\chi_{\text{SE}}\nu.
\]
It would be interesting to perform numerical simulations in the future
to test this theory numerically.
\begin{acknowledgments}
We thank Burkhard Dünweg and Anthony Ladd for informative discussions.
A. Donev was supported in part by the Air Force Office of Scientific
Research under grant number FA9550-12-1-0356. R. Delgado-Buscalioni
and F. Balboa acknowledge funding from the Spanish government FIS2010-22047-C05
and from the Comunidad de Madrid MODELICO-CM (S2009/ESP-1691). Collaboration
between A. Donev and R. Delgado-Buscalioni was fostered at the Kavli
Institute for Theoretical Physics in Santa Barbara, California, and
supported in part by the National Science Foundation under Grant No.
NSF PHY05-51164.
\end{acknowledgments}


\end{document}